\documentclass{amsart}
\usepackage{natbib}
\usepackage{graphicx}
\usepackage{amssymb}
\begin{document}
\begin{abstract}
For a Dirac particle in an Aharonov-Bohm (AB) potential, it is
shown that the  spin interaction (SI) operator which governs the
transitions in the spin sector of the first order S-matrix is
related to one of the generators of rotation in the spin space of
the particle. This operator, which is given by the projection of
the spin operator $\mathbf{\Sigma}$ along the direction of the
total momentum of the system, and the two operators constructed
from the projections of the $\mathbf{\Sigma}$ operator along the
momentum transfer and the z-directions close the $SU(2)$
algebra.It is suggested, then, that these two directions of the
total momentum and the momentum transfer form some sort of natural
intrinsic directions in terms of which the spin dynamics of the
scattering process at first order can be formulated conveniently.
A formulation and an interpretation of the conservation of
helicity at first order using the spin projection operators along
these directions is presented .

\end{abstract}
\title[The Spin Interaction in the Aharonov-Bohm Scattering ]{The Spin Interaction of a Dirac Particle in an Aharonov-Bohm Potential in First Order Scattering}
\author{A.Albeed}
\author{M.S.Shikakhwa}
\email{moody@ju.edu.jo}%
\address{Department of Physics, University of Jordan\\
11942--Amman, Jordan}%



\maketitle
\section{Introduction}
The perturbative treatments of the  Aharonov-Bohm (AB) effect
\citep{ahar59} received much interest in the literature as a
result of the observation \citep{feinberg,corinaldesi}that
perturbation theory fails when applied to this theory: The first
order scattering amplitude was found to differ from the exact one
when the latter is expanded to this order, and the second order
was shown to diverge. Many approaches were suggested in the
literature to remedy this problem for non-relativistic particles,
examples of which are
\cite{hagen1,boz1,ouvry,renormalization,ruijinaars,Rumiyyeh,lozano,hagen2,fainberg}.
When the spin degree of freedom is taken into account
\citep{hagenspin}, it was shown that perturbation theory works for
both non-relativistic \citep{Rumiyyeh} and relativistic particles
\citep{vera,boz2}; thanks to the  spin-magnetic moment
interaction.\\
 In a recent work \citep{shikakhwa2},  some algebraic properties of the interaction Hamiltonian
 in first order perturbation theory of a Dirac particles in an AB potential were reported
  and the general structure of the transition matrix at this order based
 on these algebraic properties was discussed. In this work, the
 spin interaction (SI) ( the interaction that remains after integrating out the
 space degrees of freedom ) in the first order S-matrix of a Dirac particle in the AB
 potential is investigated. It is shown that this SI is related to the generators
 of rotation in the spin space of the particle.It is suggested that the spin dynamics
 at this order can be formulated in terms of the projections of the spin operators
 along two directions characteristic of the scattering process; the total momentum and the
 momentum transfer directions. Conservation of helicity at this order is
 formulated using these operators.

\section{Algebra of the  Spin Interaction}
A Dirac particle in an electromagnetic field is governed by the
Hamiltonian ($\hbar=c=1$):
\begin{equation}
\label{1}H=H_0+H_{\mathrm{int}}
\end{equation}
where
\begin{equation}
\label{2}H_0=\boldsymbol{\alpha}\cdot\mathbf{p}+\beta m
\end{equation}
and
\begin{equation}
\label{3}H_{\mathrm{int}}=eA_0-e\boldsymbol{\alpha}\cdot\mathbf{A}
\end{equation}
Here, $e$ is the charge of the particle,
${\alpha}_i=\beta\gamma_i$ and $\beta=\gamma_4$. The $\gamma$'s
are the Dirac matrices: $\{\gamma_\mu,\gamma_\nu\}=2g_{\mu\nu}$.
The first order $S$-matrix element for the particle is :
\begin{equation}
\label{5}S_{fi}^{(1)}=- i\int {d^4 x\,\bar \psi _f \left( x
\right)\left( {e\gamma _\mu  A^\mu  } \right)\psi _i \left( x
\right)}.
\end{equation}
For the AB-potential \cite{ahar59}, we have:
\begin{equation}
\label{6}A_0=0,
\end{equation}
and
\begin{equation}
\label{7}\mathbf{A}=\frac{\Phi}{2\pi\rho}\hat{\epsilon}_\varphi,
\end{equation}
where $\rho=\sqrt{x^2+y^2}$, $\hat{\epsilon}_\varphi$ is the unit
vector in the $\varphi$-direction, and $\Phi$ is the flux through
the AB tube. The free Dirac wave functions are taken as:
\begin{equation}\label{7a}
\psi^{s}(x)=\frac{1}{L^{3/2}}\left(\frac{m}{E}\right)^{1/2}u(p,s)e^{-ip.x}
\end{equation}
with
\begin{equation}\label{7b}
u(p,s)=\left(\frac{E+m}{2m}\right)^{1/2}\omega(p,s)
\end{equation}
 Plugging the above vector potential into the matrix element , Eq.(\ref
{5}), carrying out the time and space integrals (taking
$\mathbf{p_i}=(p,0,0)\;,\;\mathbf{p_f}=(p\cos\theta,p\sin\theta,0)$,
$\theta $ being the scattering angle) we get the first order
$S$-matrix element as:
\begin{equation}
\label{8}S_{fi}^{(1)}=-4\pi^2\Delta |N|^{2} \delta (E_f-E_i)
u^{\dagger} _f \left( p_f,s_f \right)\left( \frac{\alpha_1 q_2
-\alpha_2 q_1}{q^2} \right)u _i \left(p_i,s_i \right).
\end{equation}
Here, $\Delta=-e\Phi/2\pi$ (in perturbative calculations
$0<\Delta<1$). $|N|^2$ is the  normalization constant appearing in
Eq.(\ref{7a}) ( times a factor of $L$ coming from the $z$-integral
), and $\mathbf{q=p_f-p_i}$ is the momentum transfer.  Since we
will be focusing on the spin interaction in this work, it is
convenient to write the above matrix element as
\begin{equation}\label{7'}
S^{(1)}_{fi}=-4\pi^2\Delta |N|^{2} \delta
(E_f-E_i)\frac{M^{(1)}}{q}
\end{equation}
with $M^{(1)}$ defined -using Dirac notation- as
\begin{equation}\label{8'}
M^{(1)}= <\mathbf{p_f};s_f|\frac{\alpha_1 q_2-\alpha_2
q_1}{q}|\mathbf{p_i};s_i>.
\end{equation}
Noting that $\alpha_i=\gamma_5\Sigma_i$, where
$\Sigma_i=\frac{i}{2}[\gamma_i,\gamma_j]\qquad ,\quad (i,j=1..3)$,
and $i\gamma_5=\gamma_1\gamma_2\gamma_3\gamma_4$, we define the
Hermitian  spin interaction operator (SI)
\begin{equation}\label{9}
K\equiv(\frac{\alpha_1 q_2-\alpha_2
q_1}{q})=\gamma_5\mathbf{\Sigma}.\mathbf{\hat{k}}
\end{equation}
with $\mathbf{\hat{k}}$
$(\mathbf{\hat{k}}=\frac{\mathbf{p_f}+\mathbf{p_i}}
{|\mathbf{p_f}+\mathbf{p_i}|}=(\cos\frac{\theta}{2},\sin\frac{\theta}{2},0)$),
being a unit vector  bisecting the scattering angle $\theta$ and
lying along the sum of the initial and final momenta of the
particle (see Fig.(1) below). Therefore, $M^{(1)}$ reads now
\begin{equation}\label{10}
M^{(1)}=
<\mathbf{p_f};s_f|\gamma_5\mathbf{\Sigma}.\mathbf{\hat{k}}|\mathbf{p_i};s_i>.
\end{equation}
\begin{figure}[htbp]
\includegraphics[width=8cm]{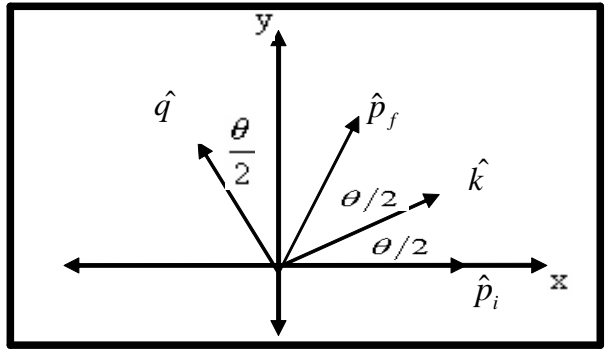}
\caption{\label{fig1} Scattering diagram in the xy-plane.}
\end{figure}
The following  algebra of the SI operator $K$ (more precisely , of
$\gamma_5K$) can be easily verified
($\mathbf{\hat{q}}=\frac{\mathbf{q}}{q}$):
\begin{eqnarray}\label{11}
  [\mathbf{\Sigma}.\mathbf{\hat{k}},\mathbf{\Sigma}.\mathbf{\hat{q}}] &=& 2i\Sigma_3 \nonumber\\
\left[\Sigma_3,\mathbf{\Sigma}.\mathbf{\hat{k}}\right] &=& 2i\mathbf{\Sigma.\hat{q}}\ \\
  \left[\mathbf{\Sigma}.\mathbf{\hat{q}},\Sigma_3\right] &=&
  2i\mathbf{\Sigma.\hat{k}},\nonumber
\end{eqnarray}
and,
\begin{equation}\label{12}
\left\{\mathbf{\Sigma}.\mathbf{\hat{k}},\mathbf{\Sigma}.\mathbf{\hat{q}}\right\}=
\left\{\Sigma_3,\mathbf{\Sigma}.\mathbf{\hat{k}}\right\}=\left\{\mathbf{\Sigma}.\mathbf{\hat{q}},\Sigma_3\right\}=0
\end{equation}
Moreover, one can easily verify that
\begin{equation}\label{13}
(\mathbf{\Sigma.\hat{k}})^2=(\mathbf{\Sigma.\hat{q}})^2=(\Sigma_3)^2=I.
\end{equation}
The above algebra, which is just the $SU(2)$ algebra says that the
two operators $\mathbf{\Sigma.\hat{k}}$ and
$\mathbf{\Sigma.\hat{q}}$ along with $\Sigma_3$ are the generators
of rotation in the spin space of the particle. To get a further
insight into this, note first that the two unit vectors
$\mathbf{\hat{q}}$ and ${\mathbf{\hat{k}}}$- which can be defined
for any similar scattering process - are functions of the
scattering angle $\theta$ and are mutually orthogonal. They are
the analogues of the $\mathbf{\hat{r}}$ and
${\mathbf{\hat{\phi}}}$ polar unit vectors in the position space.
Recalling that $\mathbf{\hat{q}}$ and ${\mathbf{\hat{k}}}$ lie
respectively, along the momentum transfer and the total momentum
directions, then it is as if the dynamics of any scattering
process define some sort of intrinsic natural coordinates for the
process at first order. Therefore, the set of operators
$\mathbf{\Sigma.\hat{k}}$, $\mathbf{\Sigma.\hat{q}}$ and
$\Sigma_3$ close the $SU(2)$ algebra in any similar scattering
process, and are merely the generators of rotation expressed in
terms of the "intrinsic" coordinates. What is special with the AB
potential, on the other hand, is the fact that the SI operator of
this potential is related (note that
$[\gamma_5,\Sigma_i]=0\quad,\forall i=1...3$) to one of these
generators, namely, $\mathbf{\Sigma.\hat{k}}$. This fact has some
consequences regarding the spin dynamics in the transition, in
particular, the conservation of helicity, as we will show below.
To start with, recall that the AB potential conserves helicity
\citep{vera}, now express the conserved helicity operators and
states in terms of the operators $\mathbf{\Sigma.\hat{k}}$ and
$\mathbf{\Sigma.\hat{q}}$ and their eigenstates. Indeed,  the
helicity operators $\mathbf{\Sigma.\hat{p_i}}$ and
$\mathbf{\Sigma.\hat{p_f}}$ of the incident and outgoing
particles, respectively,  can be expanded as:
\begin{eqnarray}
  \label{15}
\mathbf{\Sigma.\hat{p_i}}&=&\cos\frac{\theta}{2}\mathbf{\Sigma.\hat{k}}-\sin\frac{\theta}{2}\mathbf{\Sigma.\hat{q}}\nonumber\\
\mathbf{\Sigma.\hat{p_f}}&=&\cos\frac{\theta}{2}\mathbf{\Sigma.\hat{k}}+\sin\frac{\theta}{2}\mathbf{\Sigma.\hat{q}}
\end{eqnarray}
Note  the symmetry involved in these two expressions ,which is
just a reflection of the symmetry of Fig.(2). With the above
expressions at hand, it is easy now to verify with the aid of the
algebra given by Eqs.(\ref{11}-\ref{13}) the following relations:
\begin{eqnarray}\label{17}
K\mathbf{\Sigma.\hat p_i}K&=& \mathbf{\Sigma.\hat p_f} \nonumber\\
K\mathbf{\Sigma.\hat p_f}K&=& \mathbf{\Sigma.\hat p_i}
\end{eqnarray}
 Recalling that the operator $K$ is unitary in addition to being Hermitian, the
 above relations say that the spin interaction operator is a unitary
 transformation that relates the  incident and outgoing particle's helicity operators
 so that they form a couple of unitary equivalent observables. The immediate consequence of
 this is the conservation of helicity. A detailed investigation of this issue will be
 given in the next section.
 \section {\textbf{Conservation of Helicity at First Order}}
In any helicity-conserving theory, the conservation of the
helicity in first order scattering is manifested by the vanishing
of the matrix element:
\begin{equation}\label{21a}
\langle\mp;\mathbf{ \hat{p}_f} |K|\mathbf{\hat{p}_i};\pm\rangle=0.
\end{equation}
where $|\mathbf{\hat{p}_i};\pm\rangle$
($|\mathbf{\hat{p}_f};\pm\rangle$)are the eigenstates of
$\mathbf{\Sigma.\hat{p_i}}$ ($\mathbf{\Sigma.\hat{p_f}}$)with
eigenvalues $\pm 1$. We will show now that Eqs.(\ref{17}) will
naturally lead to the result (\ref{21a}). First, put
Eqs.(\ref{17}) into the form
\begin{equation}\label{21b}
\mathbf{\Sigma.\hat p_f}K\mathbf{\Sigma.\hat p_i}=K
 \end{equation}
then have both sides acting on the incident particle spinor to
get:
\begin{equation}
\label{19}
 \mathbf{\Sigma.\hat{p_f}}K|\mathbf{\hat{p}_i};\pm\rangle=\pm K|\mathbf{\hat{p}_i};\pm\rangle
\end{equation}
The state $K|\mathbf{\hat{p}_i};\pm\rangle$ is an eigenstate of $
\mathbf{\Sigma.\hat{p_f}}$ with eigenvalues $\pm1$, i.e. SI $K$
does not flip the helicity of the initial state. This , in turn,
guarantees that Eq.(\ref{21a}) holds: Using Eq.(\ref{21b}) one
has:
\begin{eqnarray}\label{20}
  \langle\mp; \mathbf{\hat{p}_f} |K|\mathbf{\hat{p}_i};\pm\rangle &=&   \langle\mp, \mathbf{\hat{p}_f}
   |\mathbf{\Sigma.p_f}K\mathbf{\Sigma.p_i}|\mathbf{\hat{p}_i};\pm\rangle\nonumber\\
   &=& - \langle\mp; \mathbf{\hat{p}_f} |K|\mathbf{\hat{p}_i};\pm\rangle
\end{eqnarray}
where we have allowed   $\mathbf{\Sigma.\hat{p_i}}$ and
$\mathbf{\Sigma.\hat{p_f}}$ to act on their eigenstates.
Therefore,
\begin{equation}\label{21}
\langle\mp;\mathbf{ \hat{p}_f} |K|\mathbf{\hat{p}_i};\pm\rangle=0.
\end{equation}
\begin{figure}[htbp]
\includegraphics[width=8cm]{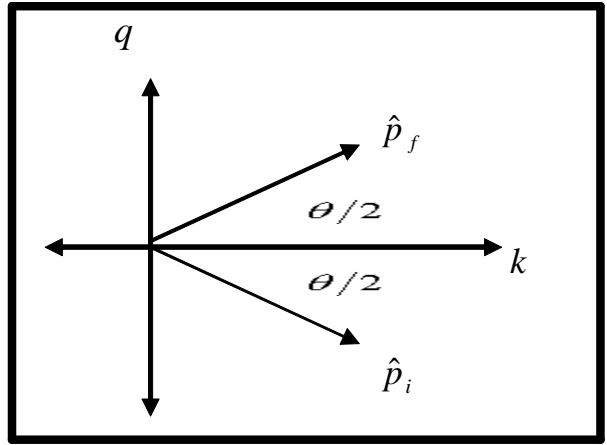}
\caption{\label{fig2} Scattering diagram in the $k-q$ plane.}
\end{figure}
 The present formulation encourages one to formulate conservation
 of helicity in the $\mathbf{\hat{k}}-\mathbf{\hat{q}}$ plane (see
 Fig.(2)). The symmetry in this figure suggests that conservation
of helicity implies the invariance of the components of both
$\mathbf{\Sigma.\hat{p_i}}$ and $\mathbf{\Sigma.\hat{p_f}}$ along
the $\mathbf{\hat{k}}$-axis, and the flipping of their components
along the $\mathbf{\hat{q}}$-axis. Indeed, the algebra given by
Eqs.(\ref{11}-\ref{13}) has the consequence
\begin{eqnarray}
  K\mathbf{\Sigma.\hat{k}}K &=& \mathbf{\Sigma.\hat{k}} \\
  K\mathbf{\Sigma.\hat{q}}K &=& -\mathbf{\Sigma.\hat{q}}
\end{eqnarray}
The above result can also be read off from the expansion of the
eigestates of $\mathbf{\Sigma.\hat{p_i}}$ and
$\mathbf{\Sigma.\hat{p_f}}$ in terms of the eigenstates of
$\mathbf{\Sigma .\hat{k}}$:
\begin{eqnarray}\label{22}
  |\mathbf{\hat{p}_i};\pm\rangle &=&  (\cos\frac{\theta}{4}\mp\sin\frac{\theta}{4}\mathbf{\Sigma.\hat{q}})|\mathbf{\hat{k}};\pm\rangle\nonumber\\
 |\mathbf{\hat{p}_f};\pm\rangle &=&
 (\cos\frac{\theta}{4}\pm\sin\frac{\theta}{4}\mathbf{\Sigma.\hat{q}})|\mathbf{\hat{k}};\pm\rangle.
  \end{eqnarray}
  The invariance of the $\mathbf{\hat{k}}$-component is established immediately:
  \begin{equation}
\langle\mathbf{\hat{k}},\pm|\mathbf{\hat{p}_i},\pm\rangle=\langle\mathbf{\hat{k}},\pm|\mathbf{\hat{p}_f},\pm\rangle=\cos\frac{\theta}{4}
\end{equation}
where we have used the result
\begin{equation}\label{23}
\langle\mathbf{\hat{k}},\pm|\mathbf{\Sigma.\hat{q}}|\mathbf{\hat{k}},\pm\rangle=0
\end{equation}
In a similar manner, if one expands using the $\mathbf{\Sigma
.\hat{q}}$ eigenstates, the flipping of the
$\mathbf{\hat{q}}$-component can be deduced immediately.

\section{Conclusions}
The effective spin interaction SI in the first order S-matrix of a
Dirac particle in an AB potential was expressed as
$K=\gamma_5\mathbf{\Sigma.\hat{k}}$, where
$(\mathbf{\hat{k}}=\frac{\mathbf{p_f}+\mathbf{p_i}}
{|\mathbf{p_f}+\mathbf{p_i}|}=(\cos\frac{\theta}{2},\sin\frac{\theta}{2},0)$
is a unit vector in the scattering plane that bisects the
scattering angle $\theta$. It was shown that the set of operators
$\mathbf{\Sigma.\hat{k}},\,\mathbf{\Sigma.\hat{q}}\,(\mathbf{\hat{q}}=\frac{\mathbf{q}}{q},\,\mathbf{q}=\mathbf{p_f-p_i})$
and $\Sigma_3$ close the $SU(2)$ algebra. This means that
$\mathbf{\Sigma.\hat{k}}$ and $\mathbf{\Sigma.\hat{q}}$ can be
identified - in addition to $\Sigma_3$- as furnishing a
representation of the generators of this group in the spin space
of the particle. Noting that the vectors $\mathbf{\hat{k}}$ and
$\mathbf{\hat{q}}$ lie along the total momentum and the momentum
transfer directions,respectively,then, they form some sort of
natural "intrinsic" coordinates for any similar scattering process
at first order. Therefore, it is natural to have the projections
of  $\mathbf{\Sigma}$ along these directions close $SU(2)$
algebra.  The fact that the spin interaction K  in the AB
potential is related to one of these generators was shown to make
the transition in the AB case  a unitary transformation with the
helicity operators of the incident and outgoing particles forming
a couple of unitary equivalent observables. This was shown, then,
to guarantee conservation of helicity at first order as should be.
Moreover,  A new view of the conservation of helicity in the first
order transition was suggested by expressing the helicity
operators and eigenstates in terms of their components along the
$\mathbf{\hat{k}}$ and $\mathbf{\hat{q}}$. It was shown, then,
that conservation of helicity was manifested as the invariance of
the component of $\mathbf{\Sigma}$ along the $\mathbf{k}$-axis and
the flipping of its component along the $\mathbf{q}$-axis  .\\

\textbf{Acknowledgements} The authors are indebted to professor
N.K.Pak and Dr. B.Tekin for a critical reading of an earlier
version of the manuscript.


\end{document}